\documentclass[twocolumn,prl,showpacs,preprintnumbers,floatfix,amsmath,amssymb]{revtex4}

\usepackage{graphicx}

\begin{document}

\title{Competition of synchronization domains in arrays of chaotic homoclinic systems.}
\author{I. Leyva$^{1,3}$}
\email{ileyva@ino.it}
\author{E. Allaria$^{1}$}
\author{S. Boccaletti$^{1}$}
\author{F.T. Arecchi$^{1,2}$}

\affiliation{$^1$ Istituto Nazionale di Ottica Applicata, Largo E. Fermi 6, 50125 Florence, Italy}
\affiliation{$^2$ Departement of Physics, University Of Firenze, Italy}
\affiliation{$^3$ Universidad Rey Juan Carlos. c/Tulipan  s/n. 28933 Mostoles Madrid, Spain}

\date{\today}

\begin{abstract}
We investigate the response of an open chain of bidirectionally
coupled chaotic homoclinic systems to external periodic stimuli.
When one end of the chain is driven by a periodic signal, the system
propagates  a phase synchronization state in a certain
range of coupling strengths and external frequencies. When two
simultaneous forcings are applied at different points of
the array, a rich phenomenology of stable competitive states is
observed, including temporal alternation and spatial coexistence
synchronization domains.
\end{abstract}

\pacs{PACS: 05.40.-a, 05.45.-a, 05.45.Xt}

\maketitle

Synchronization of chaos refers to a process wherein two (or many)
chaotic systems adjust a given property of their motion to a
common behavior \cite{reviews}. The emergence of
synchronized features has been investigated in nature
\cite{nature}, in controlled laboratory experiments
\cite{experiment,Allaria}, and the interest has moved toward the
characterization of synchronization phenomena in spatially
extended systems, such as large populations of coupled chaotic
units and neural networks \cite{populations}, globally or locally
 map lattices \cite{maps} and  continuous space
extended systems \cite{phasestc}.

In this Letter, we consider a one dimensional chain of sites,
each one undergoing a local homoclinic chaotic dynamics,
interacting via a bidirectional nearest neighbor coupling.
Homoclinic  chaos consists of a train of nearly identical spikes
 separated by erratic inter-spikes intervals (ISI). In phase space,
this  motion corresponds to the passage through a
saddle focus, where stable manifolds collapse and an unstable
manifold emerges, with the expansion rate larger than the
contraction one \cite{Shilnikov}. The saddle region displays a
large susceptibility to an external stimulus, therefore such a
chaotic system gets easily synchronized to a weak forcing
signal.

The ability of such systems to synchronize to an external
 forcing was demonstrated  in previous
works \cite{Allaria}, finding that it may constitute a reliable
communication channel \cite{Marino} robust against noise \cite{Zhou}.
Furthermore, homoclinic chaos can be self-synchronized by feeding
back a finite train of its own spikes, via either a delayed
feedback \cite{DSS} or a low frequency filter \cite{Meucci}; in
this latter case obtaining a bursting behavior reminiscent of the
dynamic of neurons in Central Pattern Generators
\cite{Rabinovich}.

However, when passing from a single system to an  array,
 a relevant problem emerges related to the ability of the array to respond to external
periodic perturbations localized at one end site, yielding synchronized
patterns. The issue we are addressing is relevant for biological or artificial communication networks.

For convenience, we refer to a chain of
dynamical  units, each one represented by a 6-variable system
modeling homoclinic chaos in a single mode CO$_2$ laser with
feedback \cite{homo}. The extension of the model to an array is:
\begin{eqnarray}
\dot{x}_1^i&=& k_0 x_1^i (x_2^i-1-k_1 \sin ^2 x_6^i), \nonumber  \\
\dot{x}_2^i&=& -\gamma_1 x_2^i- 2 k_0 x_1^i x_2^i+g x_3^i +x_4^i +p, \nonumber \\
\dot{x}_3^i&=& -\gamma_1 x_3^i+ g x_2^i + x_5^i +p, \nonumber \\
\dot{x}_4^i&=& -\gamma_2 x_4^i+ z x_2^i +g x_5^i +z p, \\
\dot{x}_5^i&=& -\gamma_2 x_5^i+ z x_3^i +g x_4^i +z p, \nonumber \\
\dot{x}_6^i&=& -\beta [x_6^i - b_0  + r (\frac{x_1^i }{1+\alpha x_1^i}+\epsilon(x_1^{i-1}+x_1^{i+1}-2<x_1^{i}>)) ] \nonumber
\label{equs}
\end{eqnarray}
Here  the index $i$ denotes the $i^{th}$ site position ($i=1,..,N$), and dots
denote temporal derivatives. For each site,
 $x_1$ represents the laser intensity, $x_2$ the
population inversion between the two levels resonant with the
radiation field, and $x_6$ the feedback voltage which controls the
cavity losses. The auxiliary variables $x_3,x_4$ and $x_5$ account for molecular exchanges
between the two resonant levels and the other rotational levels of
the same vibrational band. We consider identical units; as for
the parameters, their physical meaning has been already discussed 
\cite{homo}. Their values are: $k_0=28.5714$,
$k_1=4.5556$, $\gamma_1=10.0643$, $\gamma_2=1.0643$, $g=0.05$,
$p_0=0.016$, $z=10$, $\beta=0.4286$, $\alpha=32.8767$, $r=160$,
$b_0=0.1032$.

The coupling on each site is realized by adding to the $x_6$
equation a function of the intensity $(x_1)$  of the neighboring
oscillators. The term $<x_1^i>$ represents the average value of
the $x_1^i$ variable, calculated as a moving average over the whole evolution time.  The coupling strength
$\epsilon > 0$ is our control parameter. The system is integrated by  means of a standard fourth-order
Runge-Kutta method with open boundary conditions.

We first study the emergence of synchronization in the
absence of external stimuli, as the coupling strength $\epsilon$
increases. Due to the coupling, a spike on
one site induces the onset of a spike in the neighbor sites as
discussed in Ref. \cite{Leyva}.

\begin{figure}
\begin{center}
\includegraphics[width=8cm,height=6cm]{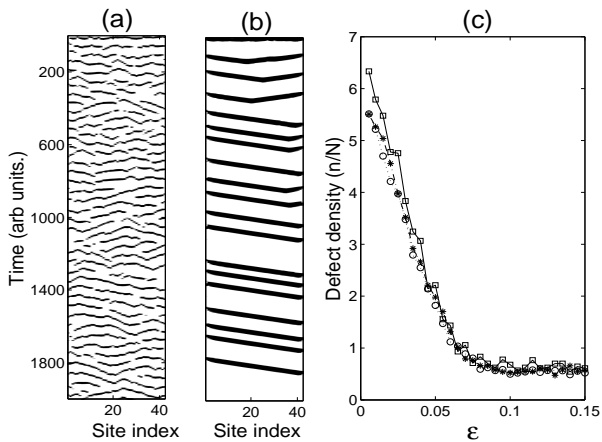}
\caption{ \footnotesize  Space-time representation of spikes positions for $\epsilon$=0.05 (a) and $\epsilon$=0.2 (b).
(c) Average defect density vs.  $\epsilon$, for different chain  lengths: N=10
($\Box$) , 40 ($\circ$), 80($\ast$). }
 \label{pos_spacetime}
\end{center}
\end{figure}

 In Fig.~\ref{pos_spacetime}(a)-(b) we show the transition from
unsynchronized to synchronized regimes by a space time
representation of the array. A detection threshold isolates the spikes
 getting rid of the chaotic small
inter-spike background, thus we plot only the spike positions as
black dots. The transition to phase synchronization is anticipated by
regimes where clusters of oscillators spike quasi-simultaneously \cite{zheng}. Clusters are
delimited by "phase slips" or defects, easily seen as
holes in the space time fabric.
More precisely, we introduce a phase measure
$\phi^i(t)$ for a time interval $t$ between two successive spikes of the same site, occurring at
$\tau_{k-1}^i$, $\tau_k^i$, by linear interpolation \cite{reviews}:
\begin{equation}
\phi^i(t)=2\pi \frac{t-\tau_{k-1}^i}{\tau_k^i-\tau_{k-1}^i}.
\end{equation}
  A defect appears as a $2\pi$ "phase slip" in the difference between the phases of two adjacent sites.
  Notice that this mutual referencing is the natural extension
of measuring the regularity of a sequence against an external clock, whenever there is no external clock,
but the time evolution  of a site compares with the nearest neighbor sites.
Hence "phase synchronization" denotes a connected line from left to right, not broken by defects.
This definition of phase synchronization does not imply equal time occurrence, thus
 the unbroken lines are not isochronous as can be seen in Fig. \ref{pos_spacetime} (b).
The cluster size increases with $\epsilon$ extending
eventually to the whole system (Fig. \ref{pos_spacetime}(b)).
The route to phase synchronization can be characterized by
the defect density, that is,  the number of defects per site.
 In Fig. \ref{pos_spacetime}(c) we plot the average defect density as function of
 $\epsilon$, calculated for a long evolution time ($T=3 \times 10^5$).
 Full phase synchronization is reached when the
defect density falls below one  defect per site.
The defect statistics has been studied for several chain lengths;  we find that above $N$=30 there are no
appreciable size-dependent effects.
Once phase synchronization is established, a further increasing of
$\epsilon$ reduces the natural frequency of the system $\omega_o(\epsilon)=\frac{2\pi}{<ISI>(\epsilon)}$.
We will show that this slowing down affects the capability of the
array to synchronize to an external signal.
 \begin{figure}
 \begin{center}
\includegraphics[width=9cm,height=6cm]{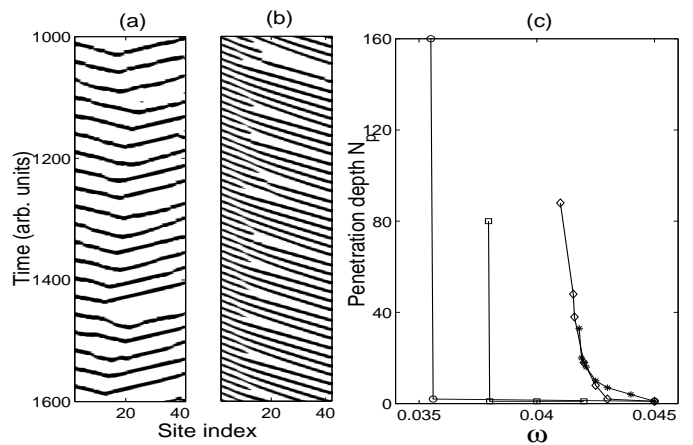}
\caption{Response of the N=40 chain with $\epsilon=0.13$  and $\omega_o(\epsilon)$=0.02 to an external periodic forcing:
  (a) $\omega$=0.015 , (b) $\omega$=0.042.
(c) Penetration depth vs. $\omega$ for different coupling strengths $\epsilon$=: 0.12 ($\ast$), 0.15 ($\diamond$),
0.2 ($\Box$), 0.25 ($\circ$).}
\label{failed_transmision}
\end{center}
\end{figure}

For this purpose, we explore the response of the system to an external
 periodic stimulus applied to the first site of the chain.
Precisely, we periodically modulate the parameter $b_o$ at the site $i=1$
as $b_{o}^1=b_o(t)=b_o(1+A\sin(\omega t))$.  From previous work, we
know that this driving can induce a phase synchronization on a
single oscillator \cite{Allaria}; here we explore the ability  of the system  to transmit the
periodic signal through the chain.

The modulation amplitude $A$ does not affect the results, 
provided that it is sufficient to
synchronize the $i=1$ site. Therefore, we do not loose
generality by fixing a constant value $A=0.3$.

We will consider that the signal has been successfully transmitted
through the system when after a finite time the last element of
the chain spikes with the same period of the external forcing,
without defects. In Fig. \ref{failed_transmision} (a)-(b) we give
examples of partial signal transmission. If  $\omega$ is too small (Fig.
\ref{failed_transmision} (a)) or too large (Fig.
\ref{failed_transmision} (b))  with respect to the natural
 frequency of the system ($\omega_o(\epsilon)$=0.02 in the figure),
 then only partial transmission is achieved.

We explore the ($\epsilon, \omega$) range over
which transmission propagates over the whole chain. In the
low frequency limit, we find that $\omega< \omega_o(\epsilon)$ is not able to globally synchronize the chain.
Independently of $N$, as  $\omega_o$ is larger than $\omega$,
the last sites tend to spike spontaneously between two consecutive
periods of the external driver  before the synchronization
propagates to them, and therefore synchronization is lost (Fig. \ref{failed_transmision} (a)).

When $\omega > 2\omega_o(\epsilon)$, the first $N_p$ sites
synchronize with the driving frequency, but beyond $N_p$  a
line of defects restores the natural oscillation regime
(Fig. \ref{failed_transmision} (b)). This
"penetration depth" $N_p$  for synchronization is invariant as
we change the whole array length. In Fig.
\ref{failed_transmision} (c) we plot the penetration depth vs. the forcing
frequency for different values of $\epsilon$.
\begin{figure}
 \begin{center}
\includegraphics[width=8cm,height=7cm]{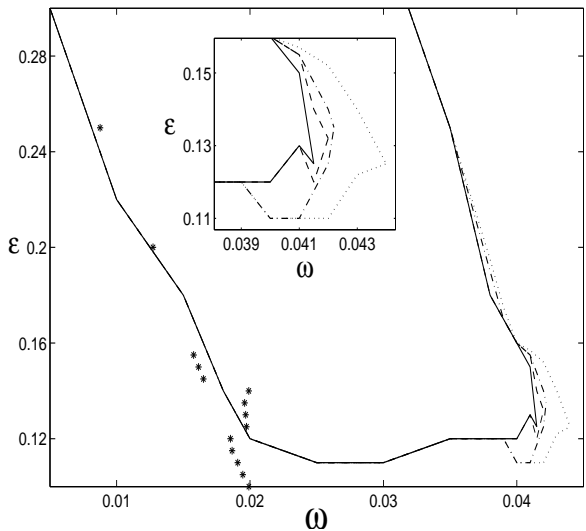}
\caption{Curves delimiting the ($\epsilon,\omega$) range for synchronized transmission in the arrays with lengths:
 N=10 (dotted lines), N=20 (dotted-dashed lines), N=40 (dashed lines) and N=80 (solid lines).
  The black stars indicate the average spiking
 frequency $\omega_o(\epsilon)$ of a site of the array in the absence of external perturbation. Inset: zoom of the area
 where the two external frequencies are selected for studying the spatial competition between synchronization domains.}
\label{arnold_range}
\end{center}
\end{figure}
If for a point ($\epsilon,\omega$)  the penetration depth is $N_p$,
 then one would observe complete synchronization only for
sizes $N<N_p$, while incomplete synchronization would
unavoidably take place for $N>N_p$. As a result, for a given $N$, only a limited range of
external frequencies can be transmitted over the whole chain. 
 
 In Fig. \ref{arnold_range} we plot the boundaries of the transmission
band as a function of $\epsilon$ and $\omega$, for several chain
lengths. The region inside the curves contains all the $(\epsilon,\omega)$
points for which global transmission is allowed.
It can be seen that for each $\epsilon$, the transmission band
extends from $\omega_o$ (black stars) to approximately 2$\omega_o$.
Notice that the system starts to transmit for
coupling strengths above the ones leading to intrinsic
synchronization (approximately $\epsilon > 0.11 $).  For weaker
couplings, the presence of defects breaks the continuity, while for $\epsilon > 0.35 $
the homoclinic dynamics is destroyed.
\begin{figure}
\begin{center}
\includegraphics[width=9cm,height=8cm]{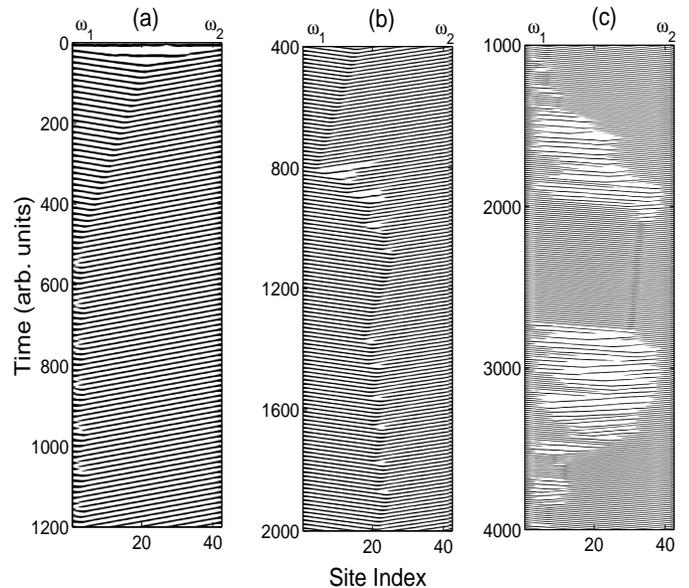}
\caption{Competition between spatial synchronization regimes induced by 
external forcing: (a)  $\omega_1=0.02,  \omega_2=0.021, \epsilon=0.13$,
 (b) $\omega_1=0.038,  \omega_2=0.042, \epsilon=0.12$,
(c) $\omega_1=0.04,  \omega_2=0.0405, \epsilon=0.11$ }
 \label{dosesti}
\end{center}
\end{figure}
The left boundary of the transmission range refers to perfect transmission
of the $\omega$ period up to the end of the chain. If one is only interested
in the transmission of the average frequency, this boundary
is slightly smeared out. 

Now we have sufficient background
to address the main question of how two different
frequencies applied at the far ends of the chain compete in
generating two separate spatial patterns of synchronization.
 The temporal competition between different
synchronization states was recently investigated theoretically
\cite{Ott} and experimentally \cite{Rajroy} in the context of a
single chaotic system forced by two external frequencies, finding
competitive behaviors as alternations of synchronism to several frequencies
($\omega_1, \omega_2$ or a combination of the two). Here we go further addressing
the problem of spatial competition between
synchronization regions, which is preliminary to
controlling the dynamics of an extended system as well as to
 studying the response of
neural assemblies to competing external perturbations.

To answer this question, we apply to the first ($i=1$) and last ($i=N$)
site two periodic perturbations with frequencies
 $\omega_1$ and $\omega_2$, respectively. For simplicity, we will take always
$\omega_o<\omega_1 < \omega_2$, so that $N_p(\omega_1) >N_p(\omega_2)$
(see Fig. \ref{failed_transmision} (c)).

The emerging competition scenario can be described with
reference to Fig. \ref{dosesti}.
For $N_p(\omega_1), N_p(\omega_2) > N$, both frequencies synchronize
over the whole chain. However, after a suitable
transient time, the whole system synchronizes to the larger
frequency $\omega_2$, with the only exception of the site $i=1$ (Fig. \ref{dosesti} (a)). This
kind of "winner-takes-all" behavior is the consequence of the extended
character of the system, and is at variance with the
single oscillator behavior for both forcing frequencies inside
the Arnold tongue. In fact, in  Ref. \cite{Ott} the entrainment takes place at
the frequency closer to the natural frequency $\omega_o$.

For $N_p(\omega_1) > N, N_p(\omega_2) < N$,  only the smaller
frequency synchronizes over the whole chain, while the larger
frequency is limited to the $N_p(\omega_2)$ sites closest to $i=N$. In
this situation, we find that permanent synchronization domains for
$\omega_1$ and $\omega_2$ are established, with an irregular domain
wall (Fig \ref{dosesti} (b)). If we increase $N$, the $\omega_2$ domain is always confined to the last N$_p(\omega_2)$
sites, independently of the total length of the chain as well as of the
value of $\omega_1$. The domains are stable, as checked with a very long integration time ($T> 3 \times10^{7}$).
\begin{figure}
\begin{center}
\includegraphics[width=8cm,height=7cm]{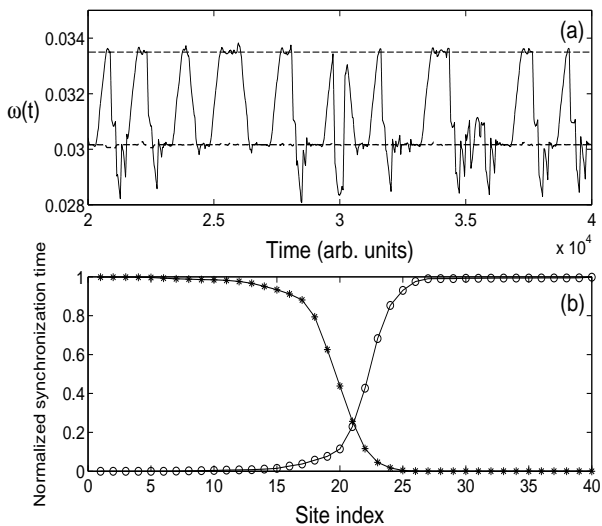}
\caption{ \footnotesize (a) Time evolution of $\omega (t)$ for
 i=1 and i=40 (dashed lines) and for the site i=22 (solid line) located on the domain boundary for the case
of Fig. \ref{dosesti} (b). (b) Normalized locking time at the frequencies  $\omega_1$ ($\ast$) and
 $ \omega_2$ ($\circ$) vs. site index for the whole chain.}
 \label{statistic}
\end{center}
\end{figure}
For the case of Fig \ref{dosesti} (b), we plot the instantaneous frequency \cite{reviews} for $i=1, i=40$ and $i=22$
(Fig. \ref{statistic} (a)). It can be observed that the site $i=22$ located on the domain boundary
locks alternatively to $\omega_1$ and $\omega_2$. The locking periods are interrupted by defects.
In Fig \ref{statistic} (b), we plot the normalized locking time of all sites
to $\omega_1$ and $\omega_2$, respectively. The transition is smooth spacewise and the boundary layer
has a width of approximately 6 sites, independently of the chain length.

Finally, for $N_p(\omega_1)$ , $N_p(\omega_2) < N $, neither of the
two frequencies stabilizes a synchronized pattern over
the whole chain. In this case, we observe alternation between
synchronization patterns with frequencies $\omega_1$ and
$\omega_2$, with intervals of asynchrony filled with defects,
as shown in Fig \ref{dosesti} (c). The duration of the
synchrony and asynchrony intervals is irregular. This competitive
behavior persists in time. As a consequence, the competition
between the two frequencies has here a cooperative
effect, insofar as it enhances the ability of each single
entrainment process to reach global synchronization over finite time slots.

In summary,  we have studied the response  of a chain of nearest
 neighbor coupled homoclinic oscillators to periodic stimuli. The array can
 propagate a synchronization state in a
range of couplings and external frequencies ($\epsilon, \omega$).
When  two simultaneous forcings are applied at different
points of the array, a rich phenomenology of stable competitive
states is observed. The features and stability of these states depend on the intrinsic
  dynamics of the system independently of the chain size.

The Authors are indebted to R. Meucci for  fruitful
discussions. Work partly supported by EU
Contract HPRN-CT-2000-00158.


\begin{thebibliography}{99}

\bibitem{reviews}
For a review of the subject see
 A. Pikovsky, M. Rosenblum and J. Kurths, {\it Synchronization: A Universal Concept in Nonlinear
Sciences}, (Cambridge University Press, 2001);
S. Boccaletti, J. Kurths, G. Osipov, D. Valladares and C. Zhou, Phys. Rep.  {\bf 366}, 1, (2002).

\bibitem{nature}
C. Schafer, M. G. Rosemblum, J. Kurths and H. H. Abel , Nature {\bf 392}, 239 (1998);
G.D. Van Wiggeren and R. Roy, Science {\bf 279}, 1198 (1998);
B. Blasius, A. Huppert and L. Stone, Nature, {\bf 399}, 354 (1999).

\bibitem{experiment}
C.M. Ticos, E. Rosa Jr., W. B. Pando, J. A. Walkenstein and M. Monti,  Phys. Rev. Lett. {\bf 85}, 2929 (2000);
D. Maza, A. Vallone, H. Mancini and S. Boccaletti, Phys. Rev. Lett. {\bf 85}, 5567 (2000).

\bibitem{Allaria}
E. Allaria,  F. T. Arecchi, A. Di Garbo and R. Meucci, Phys. Rev. Lett. {\bf 86}, 791 (2001);
S. Boccaletti, E. Allaria  R. Meucci and F. T. Arecchi. Phys. Rev. Lett. {\bf 89}, 194101 (2002) .


\bibitem{populations}
S. H. Strogatz, S.E. Mirollo and P.C. Matthews, Phys. Rev. Lett. {\bf 68}, 2730 (1992);
V. N. Belykh, I. Belykh and M. Hasler, Phys. Rev. E {\bf 63}, 036216 (2001).

\bibitem{maps}
V. N. Belykh and E. Mosekilde, Phys. Rev. {\bf E54}, 3196 (1996);
A. Pikovsky, O. Popovich and Yu. Maistrenko, Phys. Rev. Lett. {\bf 87}, 044102 (2001).

\bibitem{phasestc}
S. Boccaletti, J. Bragard,  F. T. Arecchi and H. Mancini, Phys. Rev. Lett. {\bf 83}, 536 (1999);
L. Junge and U. Parlitz, Phys. Rev.  {\bf E 62}, 438, (2000).

\bibitem{Shilnikov}
L. P. Sil'nikov, Sov. Math. Dokl. {\bf 6}, 163 (1965).

\bibitem{Marino}
I.P. Mari\~no, E. Allaria, R. Meucci, S. Boccaletti and F.T. Arecchi. Chaos (in press, March 2003).

\bibitem{Zhou}
C.S. Zhou, J. Kurths, E. Alaria, S. Boccaletti, R. Meucci and F.T. Arecchi, Phys. Rev. E {\bf 67}, 15205 (2003).

\bibitem{DSS}
F.T. Arecchi, R. Meucci, E. Allaria, A. Di Garbo and L.S. T Simring, Phys. Rev. E {\bf 67}, 46237 (2002).

\bibitem{Meucci}
R. Meucci, A. Di Garbo, E. Allaria and F. T. Arecchi, Phys. Rev. Lett.  {\bf 88}, 144101 (2002).

\bibitem{Rabinovich}
R. C. Elson, A. I. Selverston, R. Huerta, N. F. Rulkov, M. I. Rabinovich, and H.D. I. Abarbanel, Phys. Rev. Lett {\bf 81}, 5692 (1998).

\bibitem{homo}
A.N. Pisarchik, R. Meucci and F.T. Arecchi, Eur. Phys. J. D {\bf 13}, 385 (2001).

\bibitem{Leyva}
I. Leyva, E. Allaria, S. Boccaletti and F. T. Arecchi,  submitted to Phys. Rev. E (e-print: nlin.PS/0210042).

\bibitem{zheng}
 Z. Zheng, G. Huand and B. Hu, Phys. Rev. Lett. {\bf 81}, 5318 (1998)

\bibitem{Ott}
R. Breham and E. Ott, Phys. Rev E {\bf 65} 056219 (2002).

\bibitem{Rajroy}
R. Allister, R. Meucci, D. DeShazer and R. Roy, Phys. Rev E {\bf 65} 015202(R) (2003).

\end{thebibliography}
\end{document}